\documentclass[11]{article}
\usepackage[utf8]{inputenc}

\usepackage{booktabs} 
\usepackage{graphicx}
\usepackage{bbm}
\usepackage{amsmath}
\usepackage{mathrsfs}
\usepackage{color,soul}
\usepackage{geometry}
\usepackage{algorithm}
\usepackage{algpseudocode}

\usepackage{hyperref} 
\usepackage{amsmath,amssymb}
\usepackage{amsthm}

\usepackage{tikz}

\usepackage{graphicx}
\usepackage{bbm}
\usepackage{amsmath}
\usepackage{mathrsfs}
\usepackage{color,soul}
\usepackage{geometry}
\usepackage{hyperref} 
\usepackage{amsmath,amssymb}

\usepackage{amsthm}

\usepackage{subfig}

\usepackage{multirow}
\usepackage{textcomp}
\usepackage{optidef}

\usepackage{algorithm}
\usepackage{algpseudocode}
\usepackage{authblk}

\begin{document}
\title{\huge{\textbf{Energy Storage Optimization for Grid Reliability}}}

\author[1]{Md Umar Hashmi}
\author[2]{Deepjyoti Deka}
\author[3]{Lucas Pereira}
\author[1]{Ana Bu\v{s}i\'c}
\affil[1]{{INRIA and DI ENS, ENS,}
	\textit{CNRS, PSL University}, Paris,
	France}
\affil[2]{Theoretical Division,	Los Alamos National Laboratory, Los Alamos, United States}
\affil[3]{ITI, LARSyS, T\'enico Lisboa, and prsma.com}

\date{}

%
%
%

\maketitle

\begin{abstract}
	Large scale renewable energy integration is being planned for multiple power grids around the world. To achieve secure and stable grid operations, additional resources/reserves are needed to mitigate the inherent intermittency of renewable energy sources (RES). In this paper, we present formulations to understand the effect of fast storage reserves in improving grid reliability under different cost functions. Our formulations and solution schemes not only aim to minimize imbalance but also maintain state-of-charge (SoC) of storage. In particular, we show that accounting for system response due to inertia and local governor response enables a more realistic quantification of storage requirements for damping net load fluctuations. The storage requirement is significantly lower than values determined when such traditional response are not accounted for. We demonstrate the performance of our designed policies through studies using real data from the Elia TSO in Belgium and BPA agency in the USA. The numerical results enable us to benchmark the marginal effect on reliability due to increasing storage size under different system responses and associated cost functions.\\
\end{abstract}


\section{Introduction}
\begin{figure}[!htbp]
	\center
	\includegraphics[width=4in]{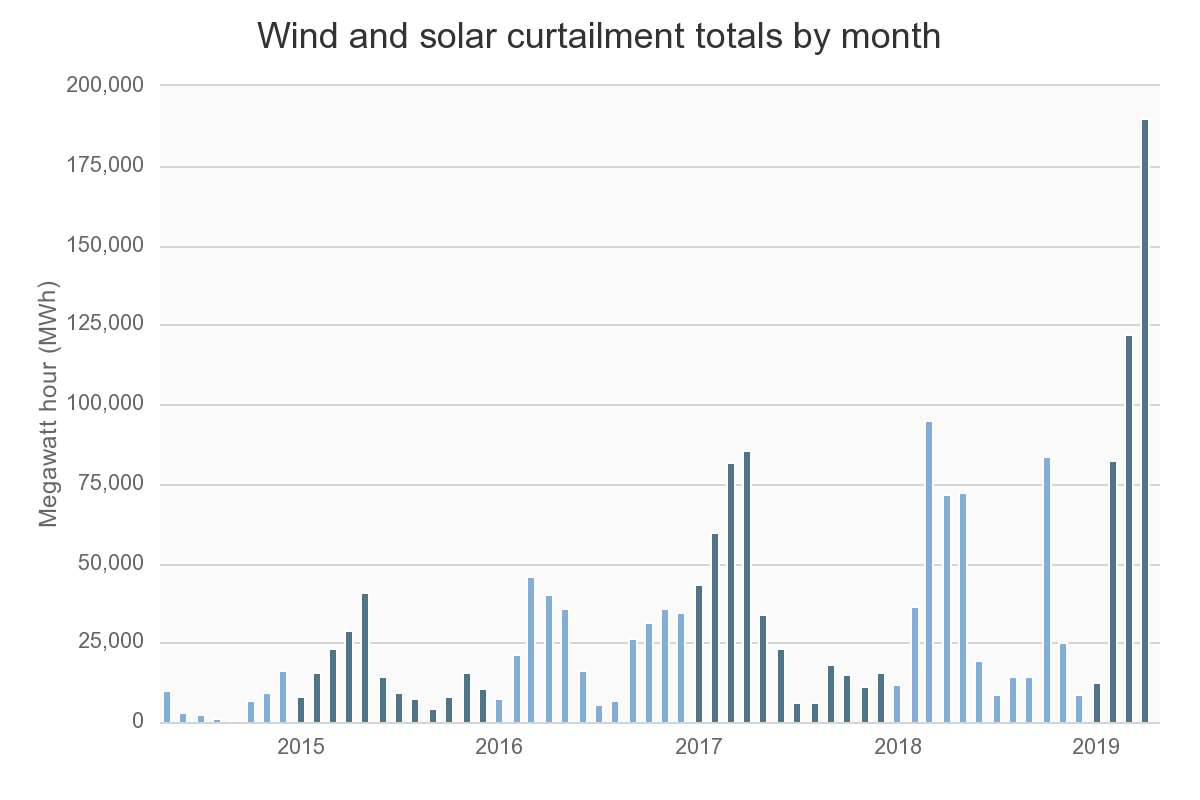} 
	\caption{CAISO (California Independent System Operator) wind and solar curtailment in MWh from 2014 to 2019.}\label{figCAISOcurtailmonthly}
\end{figure}

Massive installation of solar and wind resources in power grids is slated to replace conventional sources of power. As renewable generation is a function of weather parameters such as solar irradiance and wind speed, such sources, unlike conventional resources, are inherently uncertain/stochastic in nature. For instance, solar generation could fluctuate more than 70\% due to passing clouds during daytime and wind generation could ramp down 100\% due to loss of wind \cite{crabtree2011integrating}. As the share of renewable generation has increased, the amount curtailment has also proportionally increased, with a total cumulative curtailed energy of 1817 GWh from May 2014 to April 2019 in CAISO, see Fig.~\ref{figCAISOcurtailmonthly}\footnote{\url{http://www.caiso.com/informed/Pages/ManagingOversupply.aspx}}.

Traditional resources, due to the presence of rotation mass, provide system inertia to counter fluctuations \cite{kundur1994power}. The absence of inertia in several inverter corrected renewable resources compounds the problem of managing variability, when penetration levels increase further \cite{delille2012dynamic}. To ensure power system reliability, utilities have to hold ramp-up and ramp-down reserves in order to compensate for sudden loss of renewable generation. The authors in \cite{holttinen2008estimating} observe that, to accommodate 15\% of wind generation, traditional reserves have to be increased upto 9\%. In order to compensate the renewable volatility and avoid massive reserve procurement, additional fast ramping resources, with associated performance based payments \cite{zou2015evaluating} are being incorporated. The authors in \cite{hill2012battery} observe that energy storage systems can mitigate issues with large scale renewable integration due to their fast ramping. Falling cost of Li-Ion batteries (and other energy storage technologies) has encouraged the bulk installation of batteries for this purpose \footnote{\url{https://about.bnef.com/blog/behind-scenes-take-lithium-ion-battery-prices}}. While it is true that increasing energy storage can in theory lead to improvement in reliability, their true performance depends on available conventional responsive resources. Following a net imbalance in injection, the complete system response, due to conventional generation and fast ramping batteries, determines the dynamics in operational frequency, including the maximum deviation in frequency (termed 'nadir') and the time to reach there \cite{kundur1994power}. This is utilized by grid operators to determine reserves necessary to ensure secure operations \cite{chavez2014governor}. The overarching goal of this work is to \emph{quantify} the effect of storage, including its marginal value, on reliability by accounting for the system response in the presence of system inertia and generator governors. 
\subsection{Prior Work}
Optimization of storage operations is a growing field of research. Storage usage for arbitrage and peak shaving operates at a slower time scale (minutes-hours to weeks) and has been analyzed in \cite{mohsenian2010optimal,hashmi2019energy, bradbury2014economic,oudalov2007sizing,leadbetter2012battery,hashmi2019arbitrage}. In work associated with storage usage for reliability, the authors in \cite{akhavan2013optimal} consider investor-owned reserves that perform bidding to profitably provide balancing services \cite{hashmi2018effect}. In \cite{oudalov2007optimizing}, storage along with energy dissipating resistors are used for primary frequency control. The authors of \cite{knap2015sizing} use energy storage for providing inertial response along with primary frequency regulation and show that response similar to conventional power plants can be derived. \cite{weitemeyer2015integration} looks at the impact of energy storage parameters such as capacity, ramping parameters and conversion efficiency on the impact of renewable integration. \cite{buvsic2017distributed} observes that myopic control algorithms for storage operation can approximate deterministic solution for cases with small time-difference between decisions. \cite{su2013modeling} looks at the utility's problem of minimizing power imbalance by using storage devices and presents threshold based control rules, however they ignore system response in their analysis. \cite{das2015adequacy} shows that reserve sizing based on worst-case imbalance would not be financially plausible. Bringing the reliability requirement and system response into the picture can help in understanding the marginal increase in reliability due to integrating energy storage as reserves. 

\subsection{Contributions of the paper} 
We consider centralized optimization of utility owned/operated storage for improving grid reliability. While the profitability of battery is crucial, we assume payment for capacity to owners and do not include electricity prices while optimizing storage actions in real-time \cite{zou2015evaluating}. Rather, we are interested in identifying marginal system-wide reliability improvement due to available storage. We first justify a relaxed system reliability index measured in terms of net power balance that is principally aware of the conventional bulk system response due to inverters and governors. For both linear and quadratic cost functions of net imbalance, we present convex but non-smooth optimization formulations for real-time storage operation. The storage optimization problem can also incorporate state of charge (SoC) based constraints to ensure ramping availability during unexpected large imbalances. We present McCormick relaxation and threshold based exact schemes for our formulations, and benchmark the performance with respect to different reliability metrics, using real imbalance information from two regions: (a) Elia TSO in Belgium and (b) BPA (Bonneville Power Administration) in USA. Our work demonstrates that while increase in storage improves reliability, the \emph{marginal value of integrating storage deteriorates with the storage size}. More importantly, \emph{the computed decrease in marginal benefit is more severe for systems with higher system response.} As our optimal solutions are deterministic and require information of all fluctuations, we compare them with a myopic storage algorithm that uses only current information. We observe that due to significantly fast operation of reserves, the myopic storage operation approaches the reliability improvements in the deterministic solutions, and hence can be used for real-time operation. 

The rest of the paper is organized as follows. Section II introduces the reliability indices used in the paper and formulation of the response aware storage operation to reduce imbalance. Section III provides optimization formulation for storage as reserve for different cost function. Section IV provides case studies for Elia in Belgium and BPA in the USA respectively. Section V concludes the paper.

\section{Reliability of contemporary power grids}
In this section, we provide the definition of reliability index and a framework for our analysis. The system average interruption duration index, SAIDI \cite{accenture}, is commonly used as a reliability indicator by electric power utilities. SAIDI is the average outage duration for each customer served and is given as:
\begin{equation}
\text{SAIDI} = \frac{\text{Sum of customer interruption duration}}{\text{Total number of customers served}} =
\frac{\sum U_i N_i}{C_T}
\end{equation}
where $N_i$ is the number of consumers for the outage time $U_i$ for incident $i$ and $C_T$ is the total number of customers served. We define reliability index (RI) as
\begin{equation}
\text{RI} = \left( 1 - \frac{\text{SAIDI in units of time}}{\text{Time horizon for calculating SAIDI}} \right)\\
\label{eqrelia}
\end{equation}

\begin{table}[!htbp]
\caption {\small{Reliability of Power Network for year 2013 \cite{accenture}}}
\vspace{-10pt}
\label{reliability}
\begin{center}
\begin{tabular}{| c | c| c|}
\hline
Country & SAIDI in minutes & Reliability (\%) \\ 
\hline
China	& 480 & 99.9355358582 \\
Canada	& 311 & 99.9582326081 \\
Australia & 262 & 99.9648133226 \\
USA	& 138 & 99.9814665592 \\
Brazil	& 110 & 99.9852269675 \\
Spain & 72 & 99.9903303787 \\
UK	& 63 & 99.9915390814 \\
France	& 48 & 99.9935535858 \\
Italy & 33 & 99.9955680902 \\
Netherlands	& 17 & 99.997716895 \\
Korea	& 15 & 99.9979854956 \\
Germany & 15 & 99.9979854956 \\
Singapore	& $<$1 & $>$99.9998656997 \\
\hline
\end{tabular}
\hfill\
\end{center}
\end{table}

Almost all developed countries have a power system reliability higher than 99.9\% (see Table \ref{reliability}), that is expected to be ensured in the presence of renewables \cite{accenture}. For research purposes, the detailed real-world information for faults and consumers affected by each fault as needed for calculating SAIDI may be hard to get. Therefore, we redefine SAIDI in terms of the power imbalance in demand and supply relative to the aggregate load. We define residual $R(i) = \Delta_i + s_i$, where $\Delta_i$ and $s_i$ denote net imbalance (without storage) and storage power output at time $i$, respectively. For our system, \textit{modified} SAIDI is defined as
\begin{equation}
\text{SAIDI}^{\text{mod}} = \frac{\sum_i^{N_T} |R(i)|}{\bar{P}_g(i)} \label{saidi}
\end{equation}
where $N_T$ is the total number of samples in the time horizon for SAIDI computation. $\bar{P}_g$ is the mean of active power and is given as
\begin{equation}
\bar{P}_g =\frac{1}{N_T} \sum_{i=1}^{N_T} P_g(i).
\end{equation}
Note that the sample based SAIDI definition, similar to the cost function in \cite{su2013modeling}, intuitively assumes that the number of customers interrupted is captured in the size of power imbalance in the system. While our reliability measure increases with decreasing net imbalance, it does not account for the system response following an imbalance. 

\begin{figure}[!htbp]
	\centering
	\includegraphics[width=4.4in]{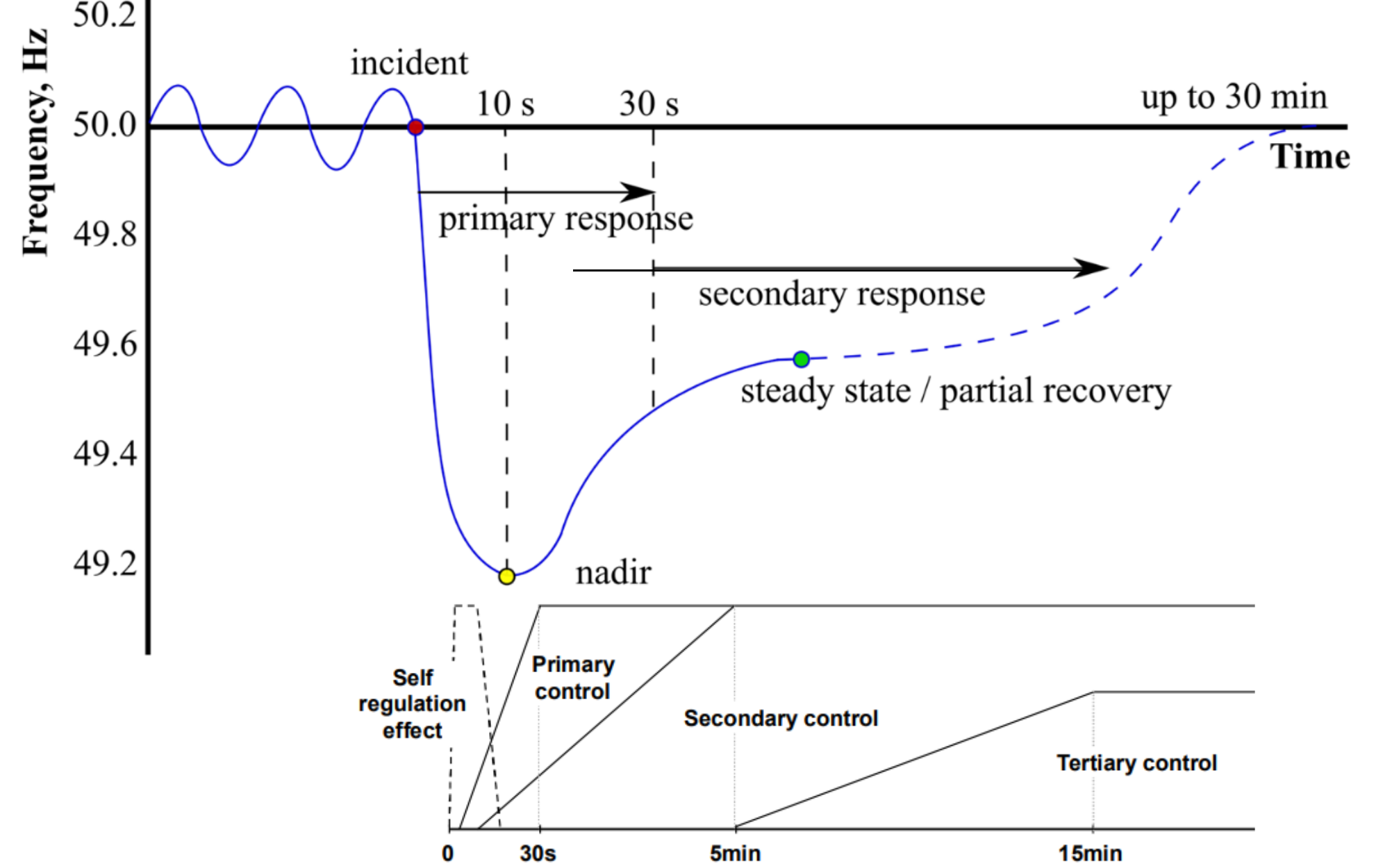}
	\caption{Reserves activation during a contingency \cite{just2015german}} 
	\label{responsereserves}
\end{figure}

\textbf{Including system response in reliability:} In power grids, demand and supply are matched approximately at every time instant to maintain frequency within a narrow band as listed in Table \ref{frequencyvar}. 
\begin{table}[!htbp]
	\caption {Continuous operating frequency range}
	\vspace{-10pt}
	\label{frequencyvar}
	\begin{center}
		\begin{tabular}{ |c| c| } 
			\hline 
			Country&COFR\\
			\hline
			Germany\cite{machado2006grid}, China \cite{hossain2014large}&	49.5 to 50.5 Hz\\
			\hline
			Australia \cite{hossain2014large} & 47 to 52 Hz\\
			\hline
			&	A-zone: 59.95 to 60.05 Hz, \\
			USA \cite{machado2006grid}	&	B-zone:59.8-59.95 \& 60.05-60.02 Hz,\\
			&	C-zone: $<$59.8 Hz \& $>$60.02Hz \\
			\hline
		\end{tabular}
	\end{center}
\end{table}
Rotational generators such as synchronous and induction machines in the grid provide an inherent rotational inertia as well as governor feedback (called Primary Frequency Control) that act to correct the operating frequency $f(t)$, following imbalance. This dynamics is modelled by the Swing equation \cite{kundur1994power,chavez2014governor}:
\begin{equation}
\frac{df(t)}{dt} = \frac{1}{M_H}(P_m(t) - P_e(t)),
\end{equation}
where $P_m$ and $P_e$ are the system's mechanical power and electrical load. $M_H$ is the inertia in the system. Considering a system-wide ramp rate of $C$ MW/s that provides frequency damping services, the frequency nadir $f_{\min}$ reached due to a net imbalance/residual $R(i)$ in the system is given by (see \cite{chavez2014governor} for the derivation):
\begin{equation}
M_HC = \frac{R(i)}{2(f_0-f_{\min}-f_{db})}. \label{fmin} 
\end{equation}
Here $f_0$ is the normal operating frequency, while $f_{db}$ is the dead-band frequency for governor response. This is derived in \cite{chavez2014governor} by first determining the time to reach the frequency nadir from the event beginning and then using that to determine the system frequency. The operation of the primary response takes places within the first $30$ seconds following an imbalance, as shown in Fig.~\ref{responsereserves}.
Note that a similar analysis can be conducted for frequency incursion above the rated frequency.

\begin{figure}[!htbp]
	\centering
	\includegraphics[width=3in]{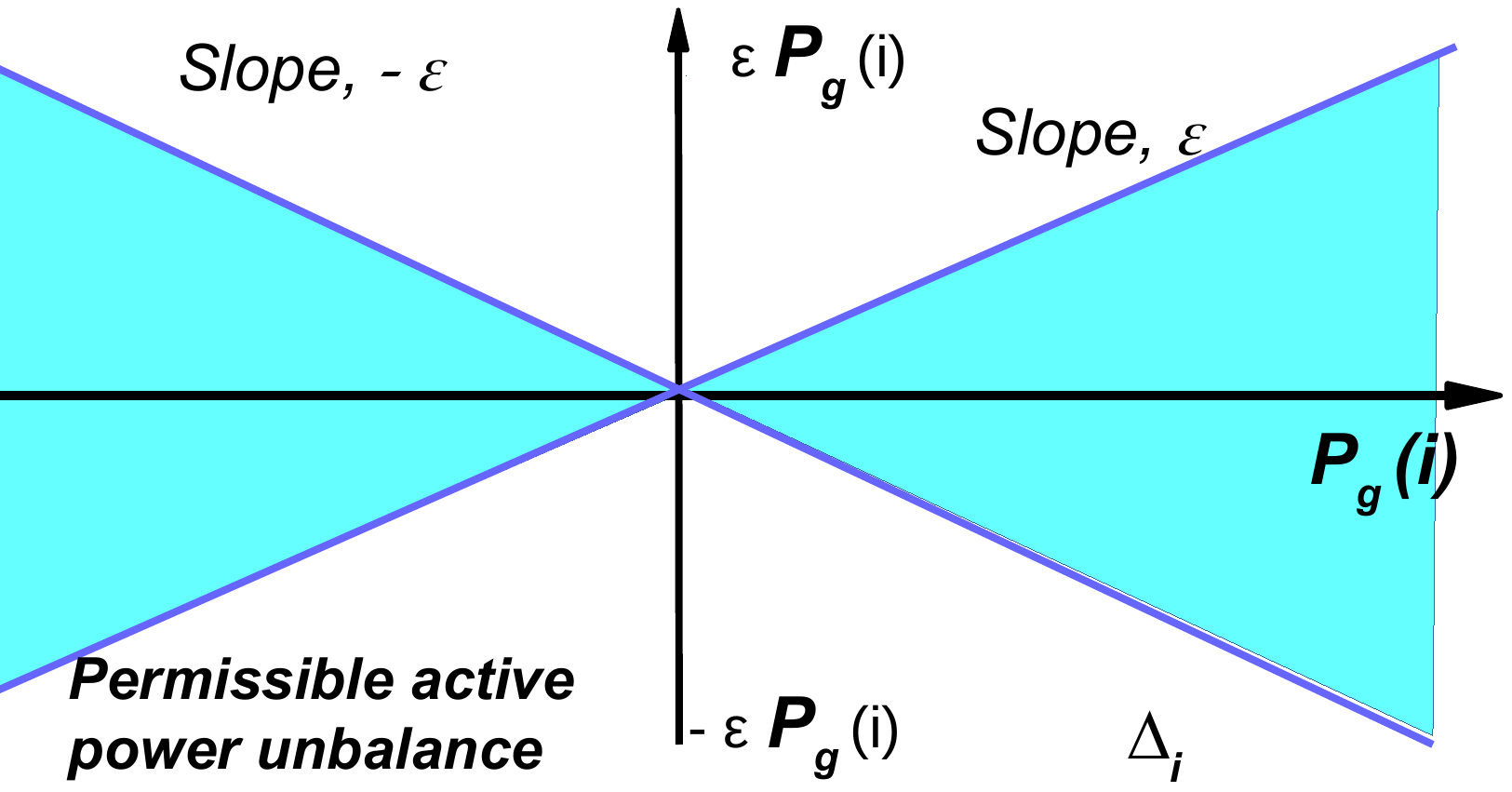}
	\caption{\small{System response based constant $\epsilon$ and permissible active power imbalance.}}\label{fluctuations}
\end{figure}

If system inertia and ramp rates of different conventional generators are comparable (else consider the minimum per mW), then the total system inertia $M_H$ and ramp rates $C$ can each be considered proportional to the total scheduled generation $P_g$  in the system. Going forward to regimes with similarly sized local generation (Eg. networked micro-grids), one can approximate $M_HC$ with a constant times $P^2_g$, the square of the total system load. Consider a pre-fixed maximum frequency deviation $f_0-f_{\min}$ for system safety. As dead-band $f_{db}$ is pre-determined, it follows from Eq.~(\ref{fmin}) that the maximum imbalance $R(i)$ that the grid can safely sustain, is proportion to the scheduled system load or generation, as noted below.
\begin{align}
-\epsilon P_g \leq R(i)=\Delta(i) + s_i \leq \epsilon P_g \label{imbalance_bound}
\end{align}
Here $\epsilon$ is a constant that depends on system inertia and ramp rate. Fig.~\ref{fluctuations} represents this permissible imbalance range, which in agreement with the observation that larger disturbances can be tolerated in system with larger load or online generation. Note that conservative operators can determine a lower $\epsilon$ to be on the safe side. 
To account for the system safety for small imbalances below $\epsilon P_g$, we modify the $\text{SAIDI}^{\text{mod}}$ formula in Eq.~(\ref{saidi}) and define response-aware $\text{SAIDI}_\epsilon^{\text{mod}}$ below.
\begin{equation}
\text{SAIDI}_\epsilon^{\text{mod}} = \frac{\sum_i \max(|R(i)| - \epsilon P_g(i), 0 )}{\bar{P}_g(i)}\label{saidimod}
\end{equation}
Using Eqs.~(\ref{eqrelia},\ref{saidi},\ref{saidimod}) the reliability for our system, with and without response awareness, are respectively given by
\begin{align}
&\text{RI}^{\text{mod}} = \left( 1 - \frac{1}{N_T}\frac{\sum_i |R(i)|}{\bar{P}_g(i)} \right)= \left( 1 - \frac{\sum_i |R(i)|}{\sum_i {P}_g(i)} \right),
\label{eqrelia2}\\
&\text{RI}_\epsilon^{\text{mod}} = \left( 1 - \frac{\sum_i \max(|R(i)| - \epsilon P_g(i), 0 )}{\sum_i {P}_g(i)} \right).
\label{eqrelia5}
\end{align}
\vspace{2mm}

\textbf{Time-scale of battery operation:} Note that we assume storage operation to be without delay after an imbalance is observed. In practice, such seamless storage operations can be conducted through rate of change of frequency (RoCoF) measurements \cite{greenwood2017frequency} directly or through the use of phasor measurement devices. In the next section, we describe the storage optimization problems that consider the defined reliability functions $\text{RI}^{\text{mod}}$ and $\text{RI}_{\epsilon}^{\text{mod}}$ (with response awareness).

\section{Optimization of Storage}
In this section, we outline optimization formulations for battery performing supply-demand balancing considering (a) linear or quadratic cost function for minimizing imbalance, (b) response of the power network, (c) maintaining the SoC of the battery. While we first consider optimal deterministic solutions schemes over a time-horizon with perfect information of fluctuations, in real-world future information will not be available. Thus, we also provide myopic rule-based real-time algorithms and benchmark them against the optimal deterministic formulations. For normalized time-instance $i$, the battery energy $b_i$ and power output $s_i$ needs to satisfy the following constraints:
\begin{align}
&b_i \in [b_{\min}, b_{\max}],\quad s_i \in [S_{\min}, S_{\max}]\label{bound_const}\\
&{b_i = b_{i-1} + \max(s_i,0)\eta_{ch} -\max(-s_i,0)/\eta_{dis}}.\label{dynamic_const} 
\end{align}
where Eq.\eqref{bound_const} reflects the bounds on $b_i$ and $s_i$, and Eq.\eqref{dynamic_const} describes the linear dynamics in $b_i$. $\eta_{ch}$ and $\eta_{dis}$ are the efficiencies of battery charging and discharging. The state-of-charge of battery at time $i$ is
\begin{equation}
\text{SoC}_i = b_{i}/b_{\text{rated}},\label{rated}
\end{equation}
where $b_{\text{rated}}$ denotes the rated battery capacity.
\subsection{Linear Cost Function}
\begin{figure}
	\centering
	\includegraphics[width=6in]{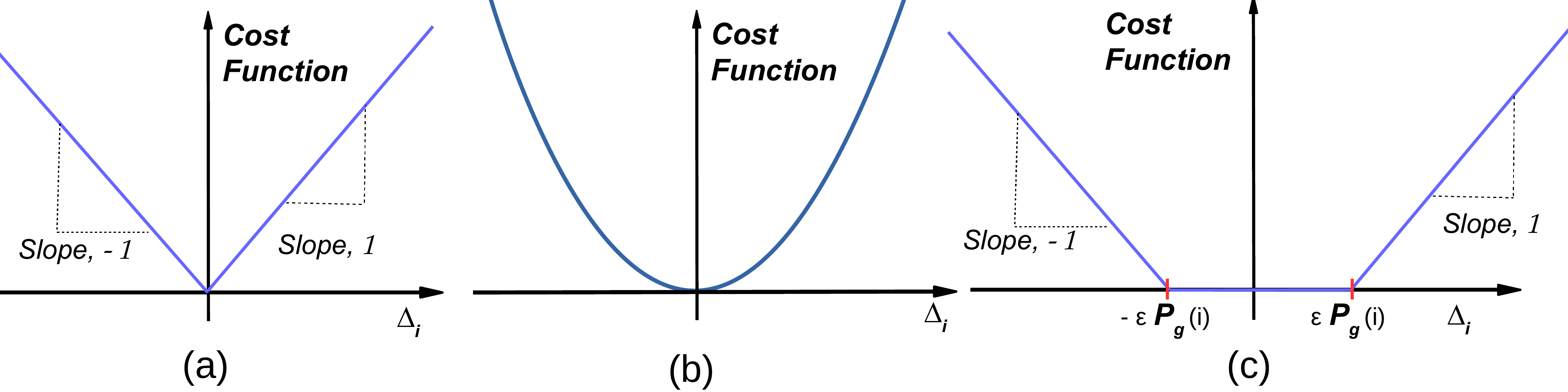}
	\caption{Cost function for imbalance minimization: (a) linear without response awareness, (b) quadratic without response awareness, (c) linear with response awareness.} \label{fig:costfunction}
\end{figure}
The linear cost functions are depicted in Fig.~\ref{fig:costfunction} (a),(b) and follow from the definition of $\text{RI}^{\text{mod}}$ and $\text{RI}_{\epsilon}^{\text{mod}}$ in Eqs.~(\ref{eqrelia2}) and (\ref{eqrelia5}) respectively. Under response awareness, the cost for imbalance below a threshold is $0$. We first describe the case where system response is not considered in making storage decisions.
\subsubsection{Reliability without response awareness}
The optimization problem $(P_{L})$ is given as
\begin{align}
&(P_L)~~~\min \sum_{i=1}^N |\Delta_i + s_i|, \quad\text{subject to} ~ (\ref{bound_const},\ref{dynamic_const})
\nonumber
\end{align}
Note that for a linear cost function, the non-zero net/marginal improvement made in reducing imbalance is the same irrespective of time-instant or overall imbalance magnitude. Thus, storage can be operated myopically considering only the current imbalance, using thresholds, as described in Algorithm~\ref{alg:linearimbalance}. \\

\begin{algorithm}[!ht]
	\small{\textbf{Inputs}: {$\Delta=(\Delta_1,\Delta_2,\ldots,\Delta_N)$, $b_0$}}, ~\textbf{Parameters}: {$ b_{\max}, b_{\min}, \delta_{\max}, \delta_{\min}, \eta_{\text{ch}}, \eta_{\text{dis}}$}\\
	\textbf{Outputs}: {$s^*$=$(s_1^*,s_2^*,..,s_N^*)$, $b^*$=$(b_1^*,b_2^*,..,b_N^*)$}\\
	\textbf{Function}: {Computes the optimal storage action for imbalance minimization under linear cost function}
	\begin{algorithmic}[1]
		\If{$\Delta_i >0$}
		$s_i^*$ = $\max\left\{- \Delta_i, \delta_{\min}h\eta_{dis}, (b_{\max} - {b_{i-1} }){\eta_{dis}} \right\} $.
		\Else
		~~ $s_i^* =\min$ $\left\{-\Delta_i, \delta_{\max}h /\eta_{ch},
		{(b_{\min} - b_{i-1})}/{\eta_{ch}}
		\right\} $.
		\EndIf
		\State Update $b_i^* = b_{i-1}^* + \max(s_i^*,0)\eta_{ch} - \max(-s_i^*,0)/\eta_{dis}$.
		\State Increment $i =i+1$.
	\end{algorithmic}
	\caption{{Linear Cost without Response }}\label{alg:linearimbalance}
\end{algorithm}
\subsubsection{Reliability with response awareness}
Based on $RI_\epsilon$ (see Eq.~(\ref{eqrelia5})), the optimization problem here is given by:
\begin{align}
(P^{\epsilon}_L)~\min\sum_{i=1}^N \{\max(|\Delta_i + s_i| - \epsilon P_g(i), 0)\},\text{subject to}~ (\ref{bound_const},\ref{dynamic_const})
\nonumber
\end{align}
Problem $(P^{\epsilon}_L)$ can also be solved by a rule-based policy given in Algorithm~\ref{alg:linearimbalanceinertia} that differs from Algorithm~\ref{alg:linearimbalance} due to $\epsilon$-induced thresholding in the cost-function. 

\begin{algorithm}[!hb]
	\small{\textbf{Inputs}: {$\Delta=(\Delta_1,\Delta_2,\ldots,\Delta_N)$, $b_0$}}, ~	\textbf{Parameters}: {$ b_{\max}, b_{\min}, \delta_{\max}, \delta_{\min}, \eta_{\text{ch}}, \eta_{\text{dis}}$}\\
	\textbf{Outputs}: {$s^*$=$(s_1^*,s_2^*,..,s_N^*)$, $b^*$=$(b_1^*,b_2^*,..,b_N^*)$}\\
	\textbf{Function}: {Computes the optimal storage action for imbalance minimization under linear cost + inertia}\\
	
	\begin{algorithmic}[1]
		\If{$\Delta_i >\epsilon P_g(i)$}
		\State $s_i^* = \max\left\{- \Delta_i + \epsilon P_g(i), \delta_{\min}h\eta_{dis}, ({b_{i-1} - b_{\max}}){\eta_{dis}} \right\} $,
		\ElsIf{$\Delta_i \in (-\epsilon P_g(i), \epsilon P_g(i))$}
		\State $s_i^* = 0$,
		\Else
		\State $s_i^* =\min$ $\left\{-\Delta_i - \epsilon P_g(i), \delta_{\max}h /\eta_{ch},
		{b_{i-1} - b_{\min}}/{\eta_{ch}}
		\right\} $.
		\EndIf
		\State Update $b_i^* = b_{i-1}^* + \max(s_i^*,0)\eta_{ch} - \max(-s_i^*,0)/\eta_{dis}$.
		\State Increment $i =i+1$.
	\end{algorithmic}
	\caption{{Linear Cost with Response}}\label{alg:linearimbalanceinertia}
\end{algorithm}
\subsection{Quadratic Cost Function}
We now describe optimal storage actions where the cost for imbalance is quadratic. In this setting, imbalance minimization at larger imbalances are prioritize for storage operation.
\subsubsection{Reliability without response awareness}
The optimization problem for storage without response awareness uses the cost function in Fig.~\ref{fig:costfunction} (c), and is formulated as:
\begin{align}
(P_Q)~~~\min \sum_{i=1}^N (\Delta_i + s_i)^2,\quad\text{subject to} ~~ (\ref{bound_const},\ref{dynamic_const})
\nonumber
\end{align}
Clearly, look-ahead is essential for solving $(P_{Q})$, unlike $(P_{L})$. However, standard convex optimization is sufficient to solve it as the cost function is smooth. 
\subsubsection{Reliability with response awareness}
Under knowledge of system response, we use a quadratic cost that ignores imbalances below $\epsilon P_g(i)$, similar to the linear setting. The optimization problem is given by:
\begin{align}
(P^{\epsilon}_{Q})~\min \sum_{i=1}^N \{\max(|\Delta_i + s_i| - \epsilon P_g(i), 0)\}^2,\text{subject to~} (\ref{bound_const},\ref{dynamic_const})\nonumber
\end{align}
The cost function in $(P^{\epsilon}_{Q})$ is not smooth due to the absolute value operator. We use $\theta_i$ to denote $\max(|\Delta_i + s_i| - \epsilon P_g(i), 0)$, and then derive a McCormick relaxation \cite{mccormick1976computability} scheme for the absolute value operator to solve $(P^{\epsilon}_{Q})$, as described below. 
\begin{gather*}
(P^{\epsilon}_{Q1})~~\min \sum_{i=1}^N \theta_i^2,\text{subject to}~~ (a)~(\ref{bound_const}),~~(b)~(\ref{dynamic_const}),\\
~~\text{(c) }\theta_i \geq 0,\quad\theta_i \geq 2z_i(\Delta_i+s_i) - (\Delta_i + s_i) - \epsilon P_g(i),\\
\text{(d) McCormick Constraints for }y_{i} = z_{i}(\Delta_i+s_i)\\
\begin{split}
&y_i \geq \Delta^{lb}_iz_i, \quad
y_i \geq (\Delta_i+s_i) + \Delta^{ub}_iz_i -\Delta^{ub}_i\\
&y_i \leq \Delta^{ub}_iz_i, \quad
y_i \leq (\Delta_i+s_i) + \Delta^{lb}_iz_i -\Delta^{lb}_i\\
\end{split}\\
\text{(e) Binary variable: } 2 y_i - (\Delta_i + s_i) \geq 0.
\end{gather*}
where $\Delta^{lb}_i = \Delta_i + S_{\min}$, $\Delta^{ub}_i = \Delta_i + S_{\max}$, $z_i$ denotes a binary variable which is equal to 1 when net imbalance with storage, $(\Delta_i+s_i)$ is positive. Note that the McCormick relaxation is used to approximate the values of a bilinear variable by creating a quadrilateral feasible space bounded by 4 constraints derived using the upper and lower limits of the individual variables in the bilinear variable. This form is exact when one of the variables in the bilinear form is binary \cite{mccormick1976computability}. Here $y_i= z_i(\Delta_i+s_i)$ has a binary component $z_i$.

Note that the storage SoC is not included in the optimization problems discussed till now. An operator may be interested in keeping SoC within a certain band to ensure available storage for future unforecasted large fluctuations. Next we discuss formulations where SoC targets are promoted through penalized SoC deviations.

\subsection{Reliability with response awareness and SoC management}
Consider the setting where storage SoC needs to be maintained within a band, denoted as $[\text{SoC}_{l}, \text{SoC}_{u}]$, where $\text{SoC}_{l}, \text{SoC}_{u}$ denote the lower and upper boundaries, and mean SoC level is denoted as $\bar{\text{SoC}} = 0.5 \times (\text{SoC}_{l} + \text{SoC}_{u})$.
We define $\gamma$ as $\gamma = \bar{\text{SoC}} - \text{SoC}_{l} = \text{SoC}_{u}-\bar{\text{SoC}}$.

Denote $\theta_i = \max(|\Delta_i + s_i| - \epsilon P_g(i), 0)$ and $\beta_i = \lambda \max(|\text{SoC}_i - \bar{\text{SoC}} | - \gamma, 0)$.
The objective of storage optimization under response awareness and SoC management for linear cost for imbalance is given as 
\begin{equation}
(P^{\epsilon}_{LS})~~~ \min \sum_{i=1}^N \{ \theta_i~~ + ~~\beta_i \}, \quad\text{subject to} ~~ (\ref{bound_const},\ref{dynamic_const},\ref{rated}).\nonumber
\end{equation}

On the other hand, the objective with quadratic cost for imbalance is given as 
\begin{equation}
(P^{\epsilon}_{QS})~~~ \min \sum_{i=1}^N \{ \theta_i~~ + ~~\beta_i \}^2, \quad\text{subject to} ~~ (\ref{bound_const},\ref{dynamic_const},\ref{rated}).\nonumber
\end{equation}
Note that with SoC management, the optimal solutions for both linear and quadratic cost formulations do not have an optimal myopic form. We, thus, revert to two Mccormick relaxation schemes to overcome the non-smooth parts of the cost function (one for reliability, another for SoC). The additional associated constraints for both $P^{\epsilon}_{LS}$ and $P^{\epsilon}_{QS}$ are given by:
\begin{gather*}
\text{(c) } \theta_i \geq 0, \quad \theta_i \geq 2z_i^1(\Delta_i+s_i) - (\Delta_i + s_i) - \epsilon P_g(i), \\
\text{(d) } \beta_i \geq 0, \quad \beta_i \geq 2z_i^2\text{SoC}_i - 2z_i^2 \bar{\text{SoC}} - \text{SoC}_i +\bar{\text{SoC}} - \gamma, \\
\text{(e) McCormick Constraints for } y_i^1=z_{i}^1(\Delta_i+s_i) \\
\begin{split}
&y_i^1 \geq \Delta^{lb}_iz_i^1, \quad
y_i^1 \geq (\Delta_i+s_i) + \Delta^{ub}_iz_i^1 -\Delta^{ub}_i\\
&y_i^1 \leq \Delta^{ub}_iz_i^1, \quad
y_i^1 \leq (\Delta_i+s_i) + \Delta^{lb}_iz_i^1 -\Delta^{lb}_i\\
\end{split}\\
\text{(e) McCormick Constraints for } y_i^2= z_{i}^2\text{SoC}_i \\
\begin{split}
&y_i^2 \geq \text{SoC}_{\min}z_i^2, \quad
y_i^2 \geq \text{SoC}_{i} + \text{SoC}_{\max}z_i^2 -\text{SoC}_{\max}\\
&y_i^2 \leq \text{SoC}_{\max}z_i^2, \quad
y_i^2 \leq \text{SoC}_{i} + \text{SoC}_{\min}z_i^2 -\text{SoC}_{\min}\\
\end{split}
\\
\text{(e) Binary variable: }
\begin{split}
2 y_i^1 - (\Delta_i + s_i) \geq 0,\\
2 y_i^2 - \text{SoC}_i \geq 0.
\end{split}
\end{gather*}
where $\Delta^{lb}_i = \Delta_i + S_{\min}$, $\Delta^{ub}_i = \Delta_i + S_{\max}$, $z_i^1$ denotes a binary variable which is equal to 1 when $(\Delta_i+s_i)$ is positive. $z_i^2$ denotes another binary variable which is equal to 1 when $\text{SoC}_i - \bar{\text{SoC}} \geq 0$ is positive.

\subsection{Myopic control of reserves considering SoC management and network inertia}
Storage control in the real-world will not have access to accurate information of future imbalances. In those settings, optimal solutions for problems ($P^{\epsilon}_{LS}$) and ($P^{\epsilon}_{QS}$) that require perfect information will not be practical. Instead, we propose a myopic Algorithm~\ref{alg:linearinertiasoc} in this section, for linear cost on reliability with response awareness and SoC management. Algorithm~\ref{alg:linearinertiasoc} is thus an extension of Algorithm~\ref{alg:linearimbalanceinertia}. It uses the current information (SoC and imbalance in the power network) and network response to make charge/discharge decisions to minimize the imbalance. When the imbalance is within bounds (see Eq.~\eqref{imbalance_bound}), it also attempts to keeps the SoC within the desired SoC band. 

Lines 3 to 6 decides whether the SoC is outside the target band. The SoC target band is decided based on battery type. For example, LiIon battery cannot be over-charged above an SoC level or over-discharged below a certain level \cite{hashmi2019optimization}.
Similarly, the zones for imbalance is identified in Algorithm~\ref{alg:linearinertiasoc}'s lines 7 to 11.

The storage operation is further constrained by capacity and ramping constraint. Based on the SoC and imbalance levels designated by $\text{Flag}_{\text{SoC}}$ and $\text{Flag}_{\Delta}$ respectively, different combinations are possible. The respective actions under each case are described in lines 12-34 of the pseudo code.

The algorithm can be similarly extended to derive a sub-optimal myopic policy for quadratic costs. In the next section, we provide simulation results on benefits from storage usage in reliability using real power grid imbalance data.

\begin{algorithm}
	\small{\textbf{Inputs}: {$\Delta=(\Delta_1,\Delta_2,\ldots,\Delta_N)$, $b_0$}}, \\	\textbf{Parameters}: {$ b_{\max}, b_{\min}, \delta_{\max}, \delta_{\min}, \eta_{\text{ch}}, \eta_{\text{dis}}$}\\
	\small{\textbf{Initialize}: $\text{SoC}_u, \text{SoC}_l, b_{\text{rated}}$}\\
	\textbf{Outputs}: {$s^*$=$(s_1^*,s_2^*,..,s_N^*)$, $b^*$=$(b_1^*,b_2^*,..,b_N^*)$}\\
	\textbf{Function}: {Computes the optimal storage action for imbalance minimization under linear cost function}
	\begin{algorithmic}[1]
		
		\State Calculate $\text{SoC}_i = b_i/b_{\text{rated}}$,
		\State Calculate $\bar{\text{SoC}} = 0.5 \times (\text{SoC}_{l} + \text{SoC}_{u})$
		
		\If{$\text{SoC}_i \leq \text{SoC}_{l}$} ~~
		$\text{Flag}_{\text{SoC}} =1$,
		\ElsIf{$\text{SoC}_i > \text{SoC}_{l} \text{ and }\text{SoC}_i \leq \text{SoC}_{u}$} ~~
		$\text{Flag}_{\text{SoC}} =2$,
		\Else ~~
		$\text{Flag}_{\text{SoC}} =3$,
		\EndIf
		\vspace{2mm}
		
		\State $\Delta_{\min} = -\epsilon P_g(i)$, $\Delta_{\max} = -\epsilon P_g(i)$
		
		\If{$\Delta_i \leq \Delta_{\min}$} ~~
		$\text{Flag}_{\Delta} =1$,
		\ElsIf{$\Delta_i > \Delta_{\min} \text{ and }\Delta_i \leq \Delta_{\max}$} ~~
		$\text{Flag}_{\Delta} =2$,
		\Else ~~
		$\text{Flag}_{\Delta} =3$,
		\EndIf
		
		\vspace{4mm}
		
		\If{ $\text{Flag}_{\text{SoC}} == 1$ and $\text{Flag}_{\Delta} == 1$ }
		 Charge, $s_i^* =$ \\ $ \max\{ \min\left\{\delta_{\max}h /\eta_{ch}, 
		(\text{SoC}_{u} - \text{SoC}_i) b_{\text{rated}}/\eta_{ch}, - \Delta_i - \epsilon P_g(i) \right\}, 0\}$,
		\vspace{3mm}
		
		\ElsIf{ $\text{Flag}_{\text{SoC}} == 1$ and $\text{Flag}_{\Delta} == 2$ }
		 Replenish charge, $s_i^* = $ \\ $ \max\{ \min\left\{\delta_{\max}h /\eta_{ch},
		(\bar{\text{SoC}} - \text{SoC}_i) b_{\text{rated}}/\eta_{ch}, - \Delta_i + \epsilon P_g(i) \right\}, 0\}$,
		\vspace{3mm}
		
		\ElsIf{ $\text{Flag}_{\text{SoC}} == 1$ and $\text{Flag}_{\Delta} == 3$ }
		\State Do nothing, $s_i^* = 0$,
		\vspace{3mm}

		\ElsIf{ $\text{Flag}_{\text{SoC}} == 2$ and $\text{Flag}_{\Delta} == 1$ }
		Charge, $s_i^* = $ \\ $ \max\{ \min\left\{\delta_{\max}h /\eta_{ch},
		(\text{SoC}_{u} - \text{SoC}_i) b_{\text{rated}}/\eta_{ch}, - \Delta_i - \epsilon P_g(i) \right\}, 0\}$,
		\vspace{3mm}

		\ElsIf{ $\text{Flag}_{\text{SoC}} == 2$ and $\text{Flag}_{\Delta} == 2$ }
		\vspace{2mm}
		\If{$\text{SoC}_i \leq \bar{\text{SoC}}$}
		Replenish charge, $s_i^* = $ \\ $ \max\{ \min\left\{\delta_{\max}h /\eta_{ch},
		(\bar{\text{SoC}} - \text{SoC}_i) b_{\text{rated}}/\eta_{ch}, - \Delta_i + \epsilon P_g(i) \right\}, 0\}$,
		\vspace{2mm}
		\Else
		Replenish charge, $s_i^* = $ \\ $ \min\{ \max\left\{\delta_{\min}h \eta_{dis},
		(\bar{\text{SoC}} - \text{SoC}_i) b_{\text{rated}}\eta_{dis}, - \Delta_i - \epsilon P_g(i) \right\}, 0\}$,
		\EndIf
		\vspace{3mm}
		
		\ElsIf{ $\text{Flag}_{\text{SoC}} == 2$ and $\text{Flag}_{\Delta} == 3$ }
		Discharge, $s_i^* = $ \\ $ \min\{ \max\left\{\delta_{\min}h \eta_{dis},
		({\text{SoC}_l} - \text{SoC}_i) b_{\text{rated}}\eta_{dis}, - \Delta_i - \epsilon P_g(i) \right\}, 0\}$,
		\vspace{3mm}
		
		\ElsIf{ $\text{Flag}_{\text{SoC}} == 3$ and $\text{Flag}_{\Delta} == 1$ }
		\State Do nothing, $s_i^* = 0$,
		\vspace{3mm}
		
		\ElsIf{ $\text{Flag}_{\text{SoC}} == 3$ and $\text{Flag}_{\Delta} == 2$ }
		Replenish charge, $s_i^* = $ \\ $ \min\{ \max\left\{\delta_{\min}h \eta_{dis},
		(\bar{\text{SoC}} - \text{SoC}_i) b_{\text{rated}}\eta_{dis}, - \Delta_i - \epsilon P_g(i) \right\}, 0\}$,
		\vspace{3mm}
		
		\ElsIf{ $\text{Flag}_{\text{SoC}} == 3$ and $\text{Flag}_{\Delta} == 3$ }
		Discharge, $s_i^* = $ \\ $ \min\{ \max\left\{\delta_{\min}h \eta_{dis},
		({\text{SoC}_l} - \text{SoC}_i) b_{\text{rated}}\eta_{dis}, - \Delta_i - \epsilon P_g(i) \right\}, 0\}$,
		\EndIf 
		\vspace{4mm}

		\State Update $b_i^* = b_{i-1}^* + [s_i^*]^+\eta_{ch} - [s_i^*]^-/\eta_{dis}$.
		\State Increment $i =i+1$.
	\end{algorithmic}
	\caption{\texttt{MyopicStorageControl}: Myopic Algorithm with Linear Cost with Response, and SoC consideration}\label{alg:linearinertiasoc}
\end{algorithm}

\section{Numerical simulations for storage optimization}
To compare benefits from our optimization algorithms for different cost functions and constraints, we use the following performance indices:
\begin{itemize}
	\item Linear deviation: ($\lambda_{\text{linear}}$) equals $\sum_{i=1}^N \{\max(|\Delta_i + s_i| -$\\ $\epsilon P_g(i), 0)\} \times 100/ \bar{P}_g(i)$, 
	\item Quadratic deviation: ($\lambda_{\text{quad}}$) equals $\sum_{i=1}^N \{\max(|\Delta_i + s_i| - \epsilon P_g(i), 0)\}^2 \times 100/ (\bar{P}_g(i))^2$, 
	\item $\text{SAIDI}_\epsilon^{\text{mod}}$ and Reliability index $\text{RI}_\epsilon^{\text{mod}}$
	\item Mean SoC
\end{itemize}
First we discuss results for data from the Elia TSO in Belgium.
\subsection{Imbalance minimization in Elia, Belgium}
\begin{figure}[!htbp]
	\center
	\includegraphics[width=4.5in]{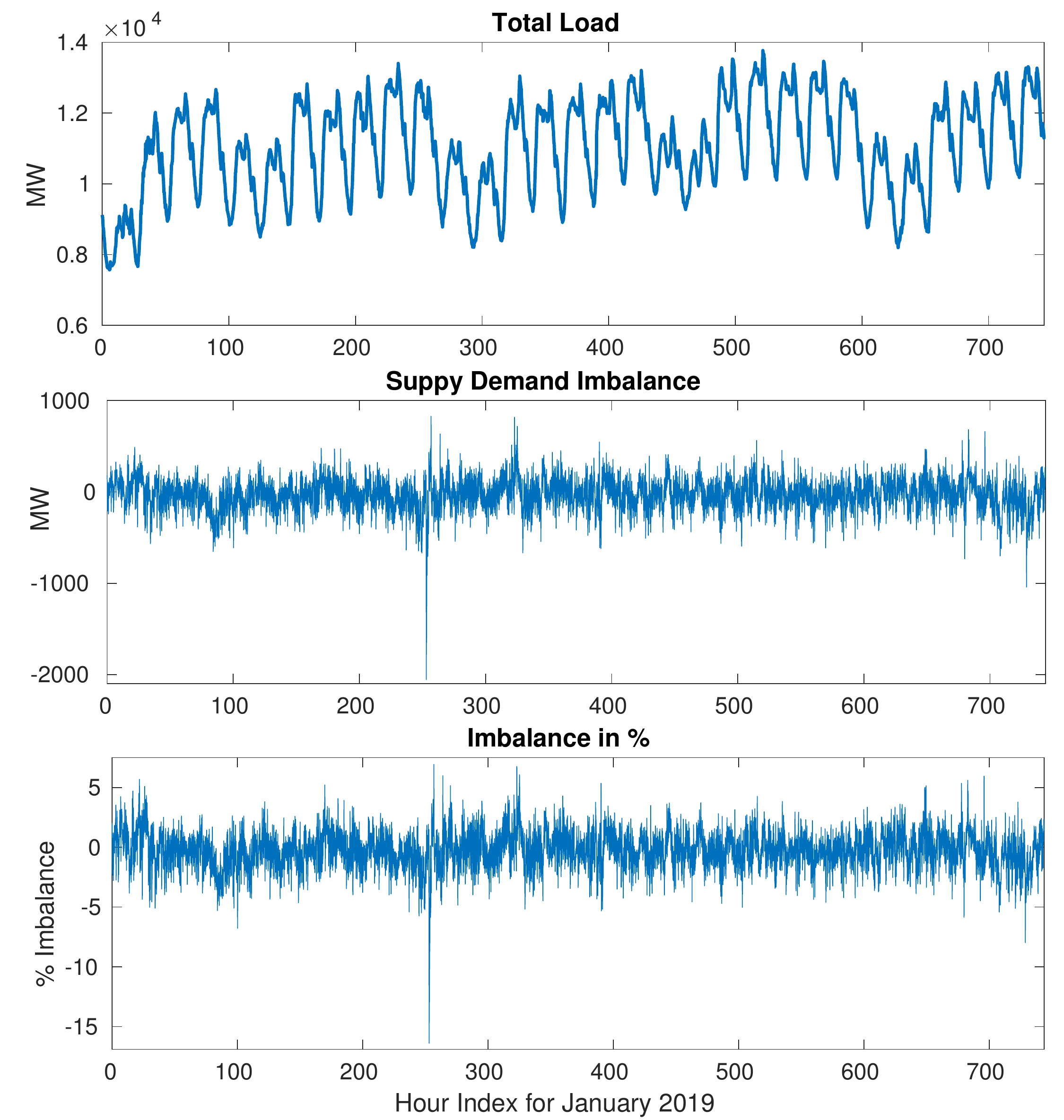}
	\caption{\small{Load, system imbalance and percentage of imbalance in Elia Belgium for the month of January 2019. Loss of a generation ($\approx 2000$MW) happening on 10th January 2019 around 13:00h; the load curve is not affected as the generation loss happened.}}\label{resfig4}
\end{figure}

The data considered in this case study is from the month of January 2019. Fig.~\ref{resfig4} shows the aggregate load, demand and supply imbalance and the imbalance in percentage with respect to the aggregate load. Without storage, the reliability $\text{RI}^{\text{mod}}$ is equal to 98.845\%. $\text{SAIDI}^{\text{mod}}$ for this month is 515.5 minutes. Observe that at hour index 253, an imbalance of the order of 17\% with respect to the total load occurs, due to a sudden loss of generation of approximately 2000 MW. The reserve sizing necessary to completely mitigate this unbalance will require an astounding ramping capability of 2000 MW or more.

The objective of study for the Elia data is to identify the marginal value of adding storage as reserves, for different values of system response, that is measured in terms of $\epsilon$ (see Eq.~(\ref{imbalance_bound},\ref{saidimod})). We vary $\epsilon$ from 0 to 5\% and implement the following 5 storage settings\footnote{
	One of the largest installed battery project is located in \href{https://en.wikipedia.org/wiki/Hornsdale_Power_Reserve}{ Hornsdale Power Reserve} in Australia. The installed capacity of this plant is 100 MW in power and 129 MWh in energy. Considering the storage installations are going to grow in future we consider significantly large battery in our numerical simulation.}: \\
(i) No storage (nominal case), \\
(ii) with 100 MW 1C-1C\footnote{Battery model xC-yC means that the battery takes 1/x hours to charge from completely discharged state at the maximum charging rate and 1/y hours to discharge from completely charged state at the maximum discharging rate},\\
(iii) with 200 MW 1C-1C,\\
(iv) with 500 MW 1C-1C,\\
(v) with 1000 MW 1C-1C.

\begin{figure}[!htbp]
	\centering
	\includegraphics[width=4in]{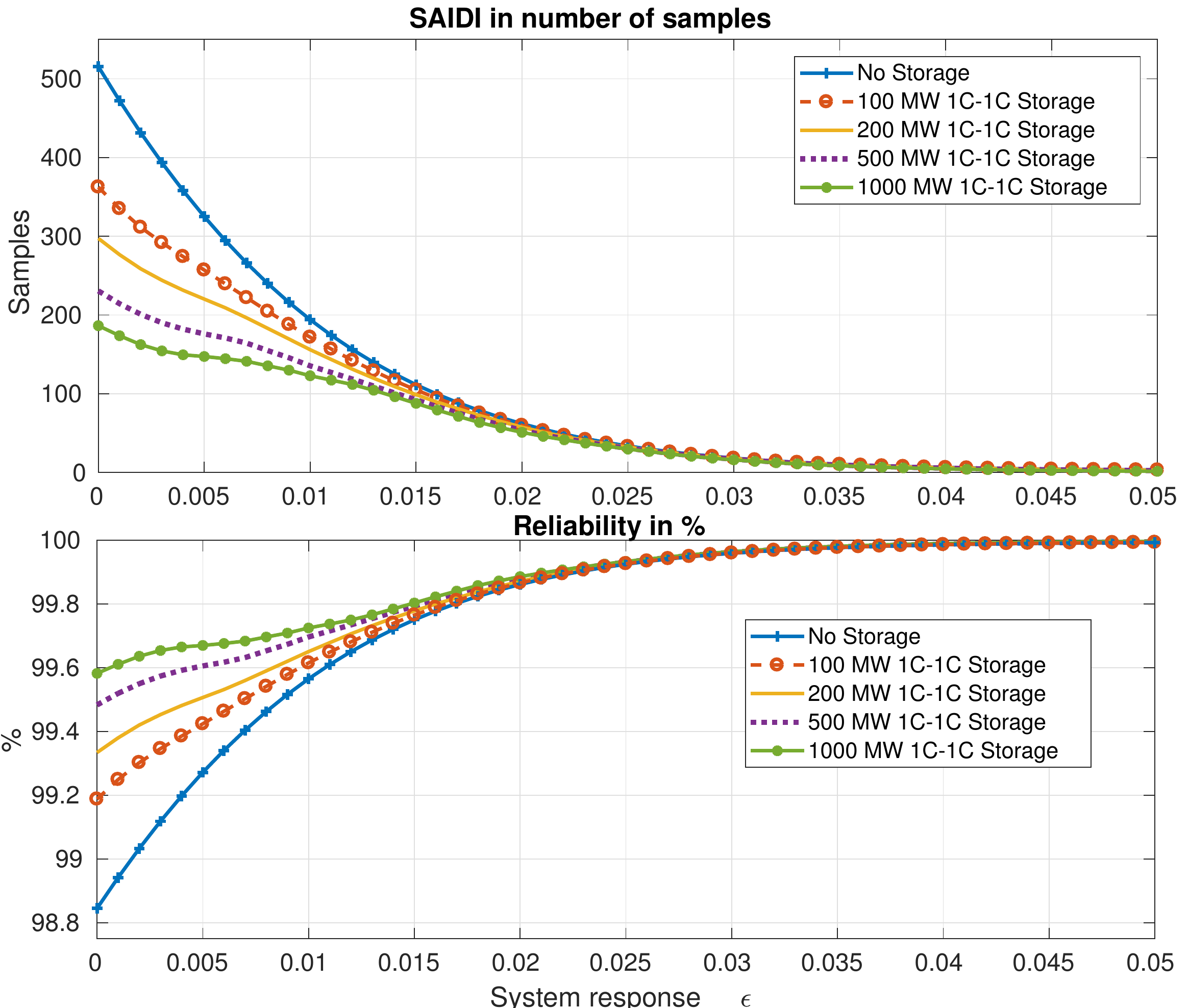}
	\hfill
	\caption{Reliability ($\text{RI}_{\epsilon}^{mod}$) and $\text{SAIDI}_{\epsilon}^{mod}$ calculated with varying system response ($\epsilon$) and storage size in Elia.}\label{figinertia}
\end{figure}

\begin{figure}[!htbp]
	\center
	\includegraphics[width=5in]{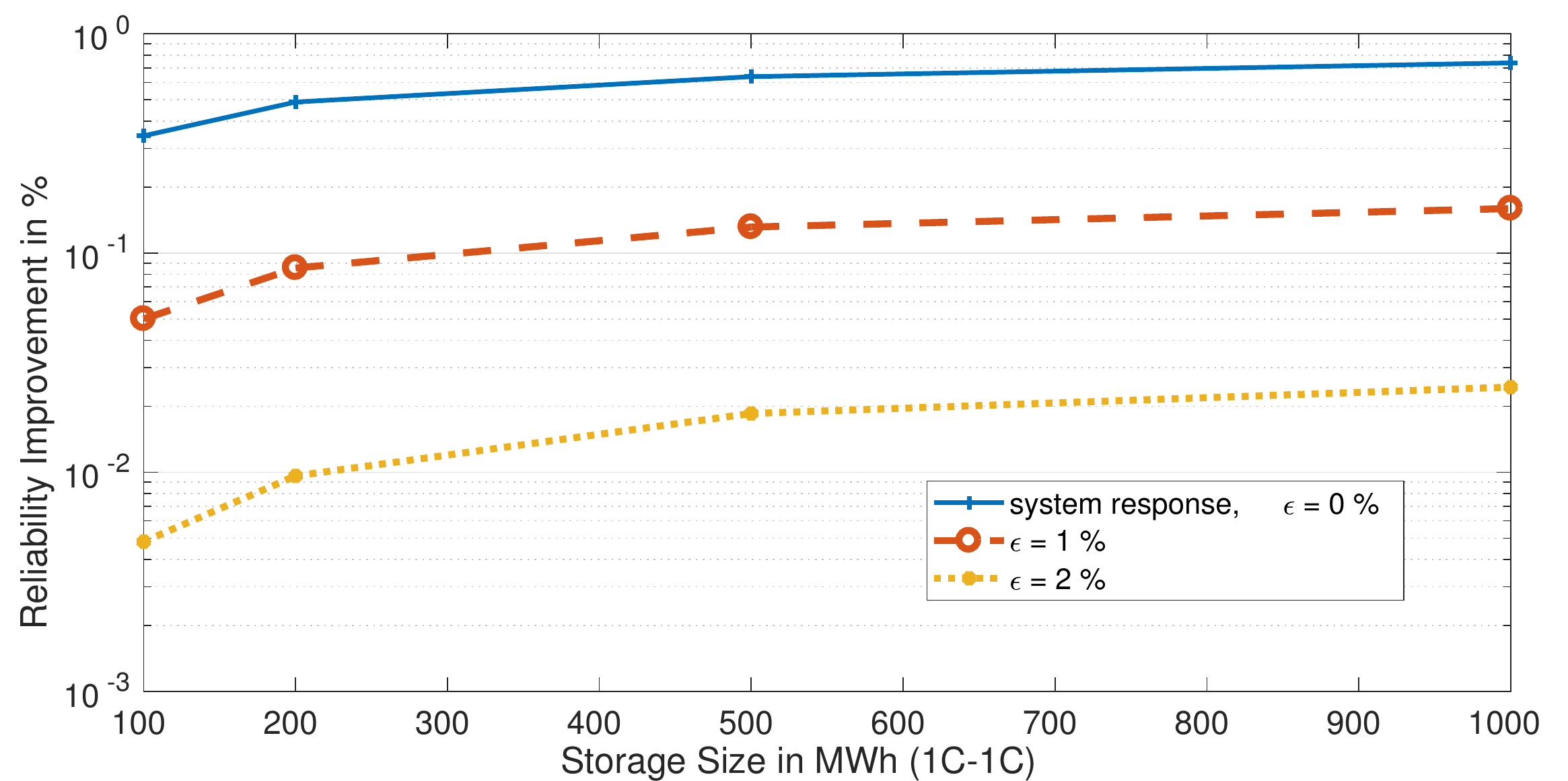}
	\caption{Marginal improvement in reliability with energy storage in Elia.}\label{figmarginalImrop}
\end{figure}

Fig.~\ref{figinertia} belabors the fact that the benefit of storage sizing for reliability is higher at lower $\epsilon$ (less conventional reserves), which is the regime of operation for grids with high renewable penetration. From Fig.~\ref{figmarginalImrop}, it is clear that the marginal benefit in reliability due to storage decreases with increasing storage sizes, as expected. However, note that the decay in marginal benefit due to increasing storage is much sharper at higher system response $\epsilon$. This suggests that analysis on greater installation of storage in a grid should involve thorough studies of current and future trends in conventional reserve availability.

\begin{figure}[!htbp]
	\center
	\includegraphics[width=5in]{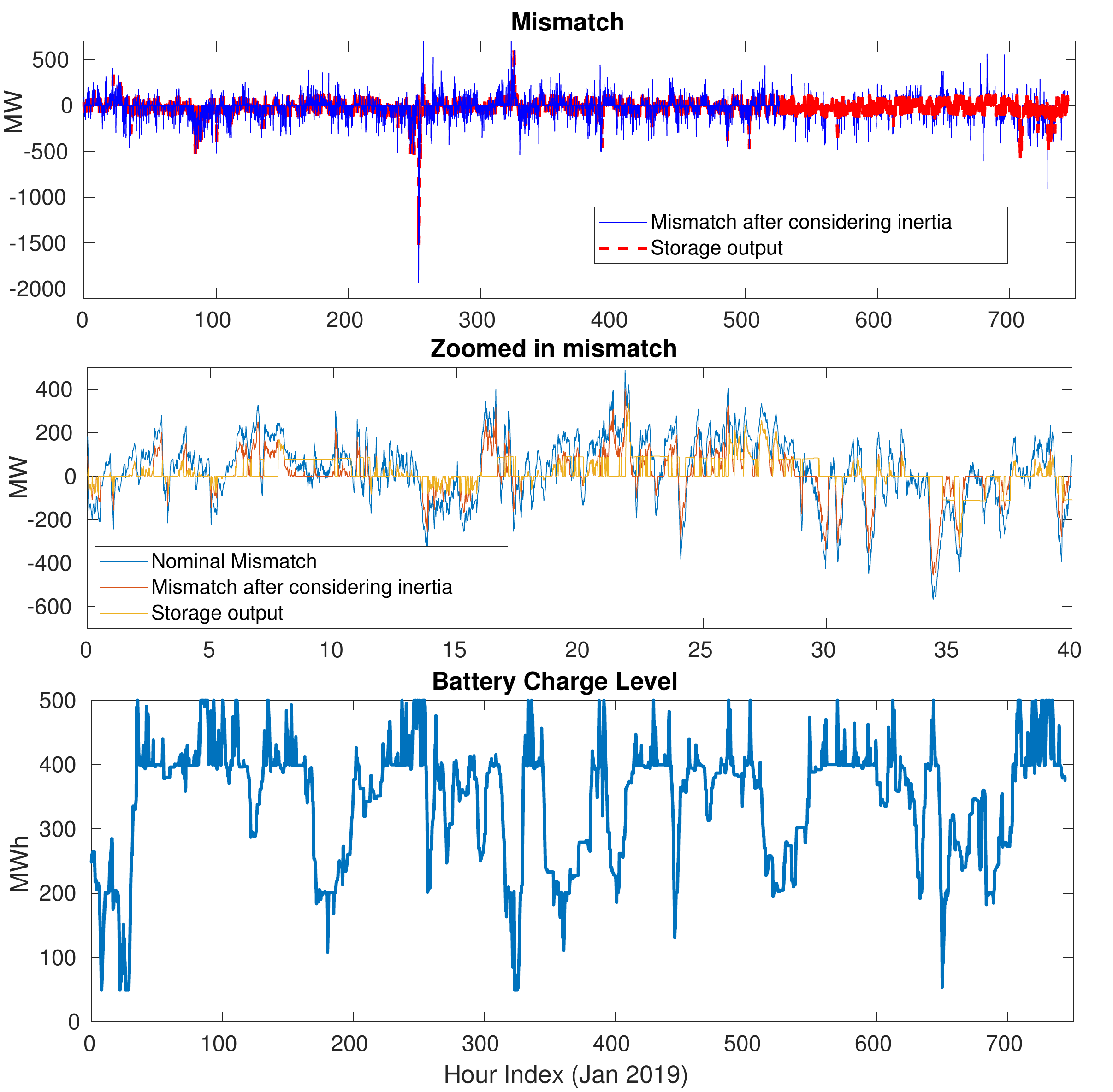} 
	\caption{\small{Myopic algorithm for storage control}}\label{figmyopic}
\end{figure}

Fig.~\ref{figmyopic} shows the battery control signal for the imbalance data in Fig.~\ref{resfig4}, using a 500 MWh (1C-1C) battery. The myopic control Algorithm~\ref{alg:linearinertiasoc} is used for the optimization with linear cost, but takes into account system response ($\epsilon = .5\%$) and SoC maintenance within $40\%-80\%$ band. Whenever the SoC dips below the minimum level or rises above the maximum, the controller follows by re-adjusting the SoC, during time instants when the imbalance $R_i$ is within the response-aware bound in Eq.~\eqref{imbalance_bound}.
\subsection{Imbalance minimization in BPA, USA}
We now consider the aggregated load and generation variation in BPA, for 6 days starting from May 10 2019, collected from their website\footnote{\url{https://transmission.bpa.gov/}}. Using this data, we compare the performance of our optimization schemes for linear and quadratic cost functions (with differing system responses), for a 1C-1C battery of capacity 100 MW. The results are provided in Table.~\ref{bpaResult}. Observe that the reliability $\text{RI}^{mod}_\epsilon$ for the myopic scheme with linear costs with response awareness and SoC, approaches that of the deterministic scheme that uses full information of all fluctuations.
Note that the $\text{RI}^{mod}_\epsilon$ for no storage case for $\epsilon = 0.05$ is 99.9197~\%. For deterministic linear cost function
$\text{RI}^{mod}_\epsilon= 99.965~\%$, for quadratic cost function $\text{RI}^{mod}_\epsilon = 99.966~\%$. For cost functions also considering SoC regulation constraint the reliability improvement for linear and quadratic deteriorates to 99.923. All the above results are for deterministic (complete information) setting. In comparison, the myopic algorithm with no look-ahead provides a reliability level of 99.943. Although, slightly lower than linear and quadratic cost functions, it is superior compared to deterministic setting with SoC regulation.
The myopic algorithm performs significantly well primarily because of the fast sampling time. A similar observation is made in \cite{buvsic2017distributed} where myopic stochastic control has an optimality gap of less than 4\% compared to the ground truth.

\begin{figure}[!htbp]
	\centering
	\includegraphics[width=4in]{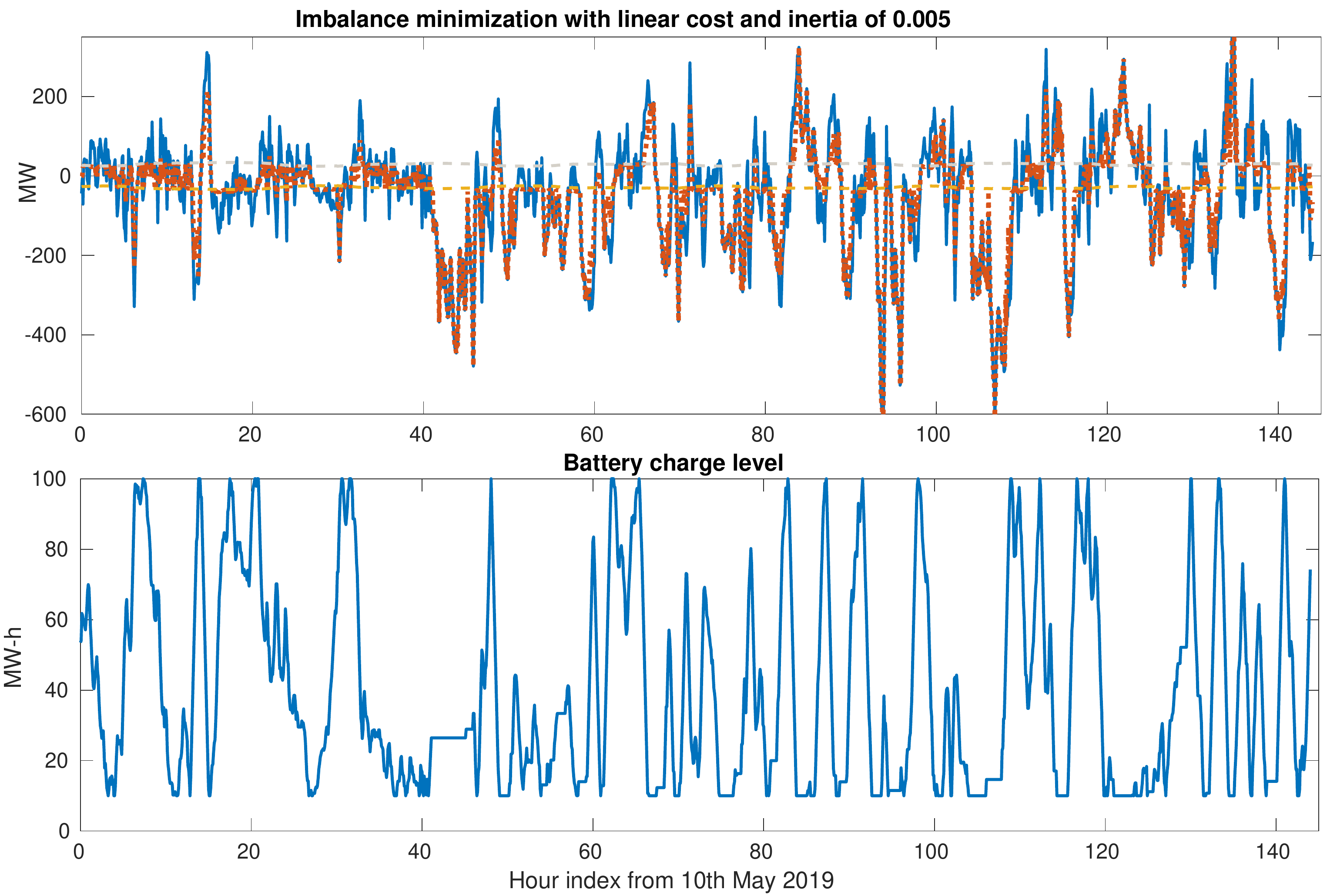}
	\caption{Imbalance tracking for linear cost with a 100 MW 1C-1C battery for system response $\epsilon = 0.005$ (no SoC maintenance) in BPA.}\label{BPAlinearInertia}
\end{figure}

\begin{figure}[!htbp]
	\centering
	\includegraphics[width=4.2in]{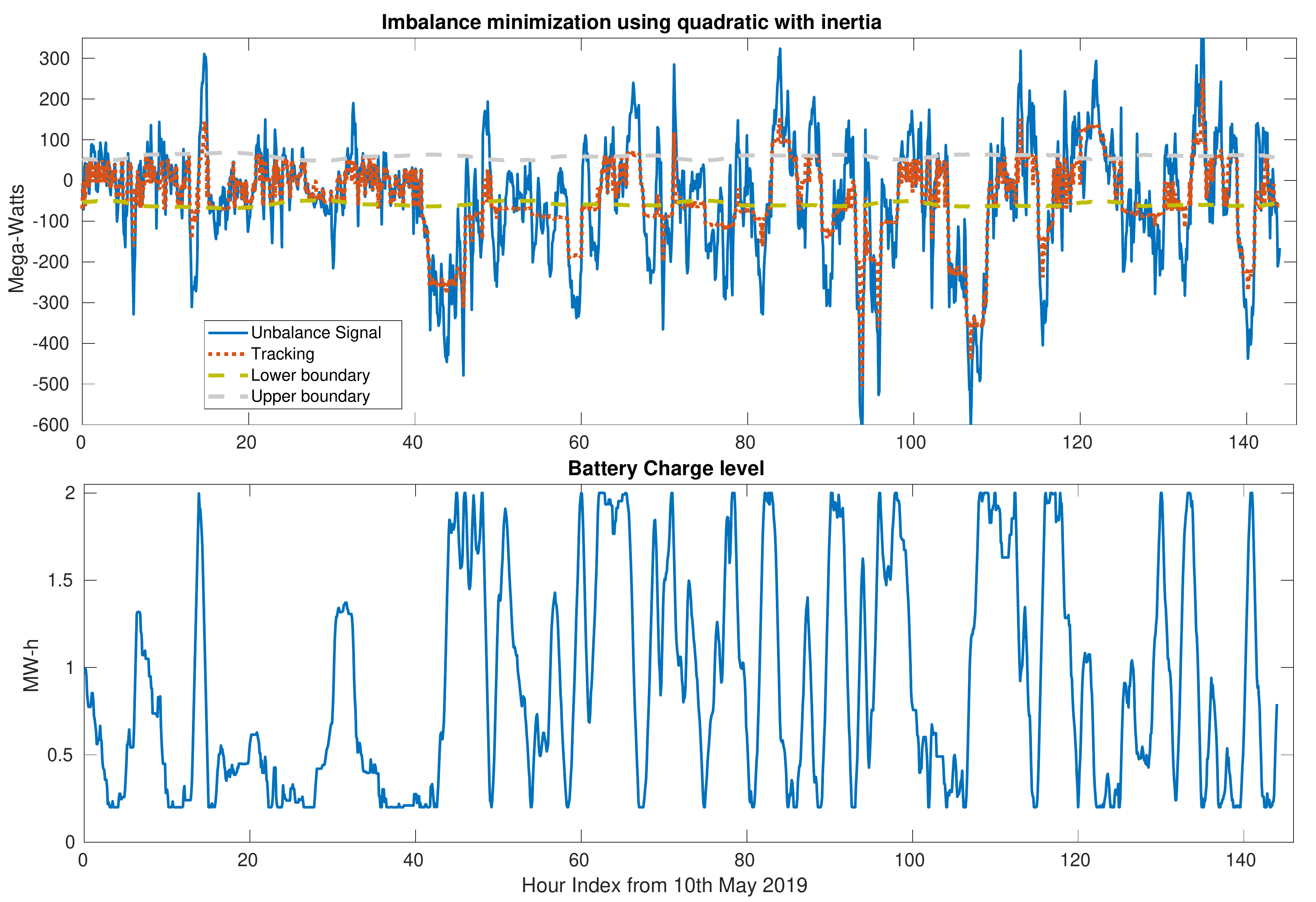} 
	\caption{Imbalance tracking for quadratic cost with a 100 MW 1C-1C battery for system response $\epsilon = 0.005$ (no SoC maintenance) in BPA.} \label{BPAquadInertia}
\end{figure}

For illustration, the tracking of imbalance signal and corresponding change in storage charge levels for response-aware linear ($P_L^\epsilon$) and quadratic ($P_Q^\epsilon$) optimization problems (no SoC) are shown in Fig.~\ref{BPAlinearInertia} and Fig.~\ref{BPAquadInertia}, respectively.
\begin{table}[!htbp]
	\caption {Performance Indices for BPA for period 10 to 15 May 2019; 1C-1C battery of capacity 100 MW.}
	\label{bpaResult}
	\begin{center}
		\begin{tabular}{ p{2.9cm}p{1.7cm}p{1.55cm}p{1.6cm}p{1.6cm}p{2.1cm}p{1.9cm} } 
			\hline 
			Optimization & $\epsilon$ & $\lambda_{\text{linear}}$ & $\lambda_{\text{quad}}$ & Mean SoC & $\text{SAIDI}_\epsilon^{mod}$ & $\text{RI}_\epsilon^{mod}$\\
			\hline
			\hline
			No storage & 0 & 3245.7	&	114.5 &-	& 32.45 & 98.1228\\
			& 0.001 & 3077.1	&	 108.1 & - & 30.77 & 98.2203 \\
			& 0.005 & 2480.7	&85.6 &-	& 24.81 & 98.5652\\
			& 0.01 & 1893.8 &63.5 &- &18.94 & 98.9047	\\
			& 0.05 & 138.8	&	3.7 & - & 1.39 & 99.9197 \vspace{0.5mm} \\ 
			\hline 
			Linear + & 0 & 1295.1& 48.1 & 0.6567 & 23.18 & 98.6592	\\
			response & 0.001 & 2168.2& 85.2 & 0.4060 &21.68 & 98.7460	\\
			& 0.005 & 1656.1& 62.2 &	0.3958 & 16.562 & 99.0421\\
			& 0.01 & 1183.9	& 41.4 &0.3939 & 11.839 & 99.3153\\
			& 0.05 &60.8	& 1.4 & 0.5426 & 0.6081 & 99.9648 \vspace{0.5mm}	\\
			\hline
			Quadratic & 0 & 1171.0	& 32.9 & 0.5222 & 23.23 & 98.6564\\
			with & 0.001 & 2165.0	& 62.9 & 0.5146 & 21.60 & 98.7505	\\
			response	& 0.005 & 1656.0& 47.5 &0.5096 & 16.52 & 99.0447 \\
			& 0.01 & 1188.6	& 33.1 &	0.4963 & 12.03 & 99.3044 \\
			& 0.05 & 57.6	& 1.1 & 0.5118 & 0.591 & 99.9658 \vspace{0.5mm}	\\
			\hline
			Linear + & 0.001 & 2175.4	& 81.8 & 0.6220 & 29.95 & 98.2681 \\
			response & 0.005 & 1689.6& 61.6 & 0.6496 & 24.01 & 98.6110 \\
			+ SoC & 0.01 & 1214.0	& 42.7 & 0.6722 & 16.02 & 99.0732 \\
			& 0.05 & 67.3	& 1.3 & 0.8191 & 1.32 & 99.9234 \vspace{0.5mm} \\
			\hline
			Quadratic & 0.001 & 2168.7	& 62.9 & 0.5210 & 22.28 & 98.7112	\\
			with & 0.005 & 1685.6& 47.6 &	 0.5590 & 19.62 & 98.8649\\
			response 	& 0.01 & 1211.9 & 33.1 &0.6063 & 15.98 & 99.0756 \\
			+ SoC	& 0.05 & 67.3	& 1.1 & 0.8171 & 1.32 & 99.9234 \vspace{0.5mm}	\\
			\hline
			Myopic & 0 & 1436.7 & 50.6 & 0.6377 & 23.18 & 98.6592 \\
			with linear & 0.001 & 1437.2 & 50.6 & 0.6394 & 22.80 & 98.6811 \\
			+ Response & 0.005 & 1440.8 &51.4 &	0.6164 & 20.35 & 98.8228\\
			+ SoC	& 0.01 & 1417.5	& 51.7 & 0.6278 & 15.818 & 99.0851	\\
			& 0.05 & 1857.1 & 58.1 & 0.6136	 & 0.9834 & 99.9430\\
			\hline
		\end{tabular}
	\end{center}
\end{table}

\section{Conclusion and Perspectives}
The paper presents algorithms for control of energy storage for minimizing demand and supply imbalance. Through theoretical motivation corroborated with numerical simulations, we show that the system dynamic response due to inertia and governor control, impacts the effect of storage in improving grid reliability. 
In particular, the marginal reliability benefit due to increasing storage decays rapidly for systems with higher conventional reserves. For real-time optimization of storage, we present myopic alternates to deterministic storage algorithms requiring full information, and show their comparable performance using real data from Elia, Belgium and BPA, USA. Furthermore, we demonstrate that storage control algorithms can maintain SoC without significant loss in reliability performance.

In future work, we plan to theoretically study the relationship between variance of stochastic imbalances, and response aware storage operation. Moreover, we plan to extend our numerical analysis to smaller grids/micro-grids with greater fraction of inverter-connected generation, and study financial incentives to maximize reliability performance from storage.

\bibliographystyle{IEEEtran}

\bibliography{ref_hard}

\begin{thebibliography}{10}
\providecommand{\url}[1]{#1}
\csname url@samestyle\endcsname
\providecommand{\newblock}{\relax}
\providecommand{\bibinfo}[2]{#2}
\providecommand{\BIBentrySTDinterwordspacing}{\spaceskip=0pt\relax}
\providecommand{\BIBentryALTinterwordstretchfactor}{4}
\providecommand{\BIBentryALTinterwordspacing}{\spaceskip=\fontdimen2\font plus
\BIBentryALTinterwordstretchfactor\fontdimen3\font minus
  \fontdimen4\font\relax}
\providecommand{\BIBforeignlanguage}[2]{{%
\expandafter\ifx\csname l@#1\endcsname\relax
\typeout{** WARNING: IEEEtran.bst: No hyphenation pattern has been}%
\typeout{** loaded for the language `#1'. Using the pattern for}%
\typeout{** the default language instead.}%
\else
\language=\csname l@#1\endcsname
\fi
#2}}
\providecommand{\BIBdecl}{\relax}
\BIBdecl

\bibitem{crabtree2011integrating}
G.~Crabtree, J.~Misewich, R.~Ambrosio, K.~Clay, P.~DeMartini, R.~James,
  M.~Lauby, V.~Mohta, J.~Moura, P.~Sauer \emph{et~al.}, ``Integrating renewable
  electricity on the grid,'' in \emph{AIP Conference proceedings}, vol. 1401,
  no.~1.\hskip 1em plus 0.5em minus 0.4em\relax AIP, 2011, pp. 387--405.

\bibitem{kundur1994power}
P.~Kundur, \emph{Power system stability and control}, vol.~7.

\bibitem{delille2012dynamic}
G.~Delille, B.~Francois, and G.~Malarange, ``Dynamic frequency control support
  by energy storage to reduce the impact of wind and solar generation on
  isolated power system's inertia,'' \emph{IEEE Transactions on Sustainable
  Energy}, vol.~3, no.~4, pp. 931--939, 2012.

\bibitem{holttinen2008estimating}
H.~Holttinen, ``Estimating the impacts of wind power on power systems—summary
  of iea wind collaboration,'' \emph{Environmental research letters}, vol.~3,
  no.~2, p. 025001, 2008.

\bibitem{zou2015evaluating}
P.~Zou, Q.~Chen, Q.~Xia, G.~He, and C.~Kang, ``Evaluating the contribution of
  energy storages to support large-scale renewable generation in joint energy
  and ancillary service markets,'' \emph{IEEE Transactions on Sustainable
  Energy}, vol.~7, no.~2, pp. 808--818, 2015.

\bibitem{hill2012battery}
C.~A. Hill, M.~C. Such, D.~Chen, J.~Gonzalez, and W.~M. Grady, ``Battery energy
  storage for enabling integration of distributed solar power generation,''
  \emph{IEEE Transactions on smart grid}, vol.~3, no.~2, pp. 850--857, 2012.

\bibitem{chavez2014governor}
H.~Ch{\'a}vez, R.~Baldick, and S.~Sharma, ``Governor rate-constrained opf for
  primary frequency control adequacy,'' \emph{IEEE Transactions on Power
  Systems}, vol.~29, no.~3, pp. 1473--1480, 2014.

\bibitem{mohsenian2010optimal}
A.-H. Mohsenian-Rad and A.~Leon-Garcia, ``Optimal residential load control with
  price prediction in real-time electricity pricing environments,'' \emph{IEEE
  Trans. Smart Grid}, vol.~1, no.~2, pp. 120--133, 2010.

\bibitem{hashmi2019energy}
M.~U. Hashmi, L.~Pereira, and A.~Busic, ``Energy storage roles in madeira,
  portugal: Co-optimizing for arbitrage, self-sufficiency, peak shaving and
  energy backup,'' in \emph{IEEE PES Powertech, Milan}, 2019.

\bibitem{bradbury2014economic}
K.~Bradbury, L.~Pratson, and D.~Pati{\~n}o-Echeverri, ``Economic viability of
  energy storage systems based on price arbitrage potential in real-time us
  electricity markets,'' \emph{Applied Energy}, vol. 114, pp. 512--519, 2014.

\bibitem{oudalov2007sizing}
A.~Oudalov, R.~Cherkaoui, and A.~Beguin, ``Sizing and optimal operation of
  battery energy storage system for peak shaving application,'' in \emph{Power
  Tech, 2007 IEEE Lausanne}.\hskip 1em plus 0.5em minus 0.4em\relax IEEE, 2007,
  pp. 621--625.

\bibitem{leadbetter2012battery}
J.~Leadbetter and L.~Swan, ``Battery storage system for residential electricity
  peak demand shaving,'' \emph{Energy and buildings}, vol.~55, pp. 685--692,
  2012.

\bibitem{hashmi2019arbitrage}
M.~U. Hashmi, D.~Deka, A.~Busic, L.~Pereira, and S.~Backhaus, ``Arbitrage with
  power factor correction using energy storage,'' \emph{IEEE Transactions on
  Power Systems}, 2020.

\bibitem{akhavan2013optimal}
H.~Akhavan-Hejazi and H.~Mohsenian-Rad, ``Optimal operation of independent
  storage systems in energy and reserve markets with high wind penetration,''
  \emph{IEEE Transactions on Smart Grid}, vol.~5, no.~2, pp. 1088--1097, 2013.

\bibitem{hashmi2018effect}
M.~U. Hashmi, D.~Muthirayan, and A.~Bu{\v{s}}i{\'c}, ``Effect of real-time
  electricity pricing on ancillary service requirements,'' in \emph{Proceedings
  of the Ninth International Conference on Future Energy Systems}.\hskip 1em
  plus 0.5em minus 0.4em\relax ACM, 2018, pp. 550--555.

\bibitem{oudalov2007optimizing}
A.~Oudalov, D.~Chartouni, and C.~Ohler, ``Optimizing a battery energy storage
  system for primary frequency control,'' \emph{IEEE Transactions on Power
  Systems}, vol.~22, no.~3, pp. 1259--1266, 2007.

\bibitem{knap2015sizing}
V.~Knap, S.~K. Chaudhary, D.-I. Stroe, M.~Swierczynski, B.-I. Craciun, and
  R.~Teodorescu, ``Sizing of an energy storage system for grid inertial
  response and primary frequency reserve,'' \emph{IEEE Transactions on Power
  Systems}, vol.~31, no.~5, pp. 3447--3456, 2015.

\bibitem{weitemeyer2015integration}
S.~Weitemeyer, D.~Kleinhans, T.~Vogt, and C.~Agert, ``Integration of renewable
  energy sources in future power systems: The role of storage,''
  \emph{Renewable Energy}, vol.~75, pp. 14--20, 2015.

\bibitem{buvsic2017distributed}
A.~Bu{\v{s}}i{\'c}, M.~U. Hashmi, and S.~Meyn, ``Distributed control of a fleet
  of batteries,'' in \emph{2017 American Control Conference (ACC)}.\hskip 1em
  plus 0.5em minus 0.4em\relax IEEE, 2017, pp. 3406--3411.

\bibitem{su2013modeling}
H.-I. Su and A.~El~Gamal, ``Modeling and analysis of the role of energy storage
  for renewable integration: Power balancing,'' \emph{IEEE Transactions on
  Power Systems}, vol.~28, no.~4, pp. 4109--4117, 2013.

\bibitem{das2015adequacy}
K.~Das, M.~Litong-Palima, P.~Maule, and P.~E. S{\o}rensen, ``Adequacy of
  operating reserves for power systems in future european wind power
  scenarios,'' in \emph{2015 IEEE Power \& Energy Society General
  Meeting}.\hskip 1em plus 0.5em minus 0.4em\relax IEEE, 2015, pp. 1--5.

\bibitem{accenture}
``Forging a path toward a digital grid global perspectives on smart grid
  opportunities,,'' Accenture’s Digitally Enabled Grid program,
  \url{https://tinyurl.com/yyvmtdod}, 2013.

\bibitem{machado2006grid}
I.~Machado and I.~Arias, ``Grid codes comparison,'' 2006.

\bibitem{hossain2014large}
J.~Hossain and A.~Mahmud, \emph{Large scale renewable power generation:
  advances in technologies for generation, transmission and storage}.\hskip 1em
  plus 0.5em minus 0.4em\relax Springer Science \& Business Media, 2014.

\bibitem{just2015german}
S.~Just, ``The german market for system reserve capacity and balancing,'' 2015.

\bibitem{greenwood2017frequency}
D.~Greenwood, K.~Y. Lim, C.~Patsios, P.~Lyons, Y.~S. Lim, and P.~Taylor,
  ``Frequency response services designed for energy storage,'' \emph{Applied
  Energy}, vol. 203, pp. 115--127, 2017.

\bibitem{mccormick1976computability}
G.~P. McCormick, ``Computability of global solutions to factorable nonconvex
  programs: Part i—convex underestimating problems,'' \emph{Mathematical
  programming}, vol.~10, no.~1, pp. 147--175, 1976.

\bibitem{hashmi2019optimization}
M.~U. Hashmi, ``Optimization and control of storage in smart grids,'' Ph.D.
  dissertation, PSL Research University, 2019.

\end{thebibliography}

\end{document}